\documentclass{ws-procs9x6}

\usepackage{epsfig}
\usepackage{amssymb}
\usepackage{feynmf}
\usepackage{axodraw}

\newcommand{\ie}{{\it i.e.}}

\newcommand{\mrm}[1]{\mbox{\rm #1}}
\newcommand{\beq}{\begin{equation}}
\newcommand{\eeq}{\end{equation}}
\newcommand{\bea}{\begin{eqnarray}}
\newcommand{\eea}{\end{eqnarray}}

\newcommand{\Eq}[1]{Eq.(\ref{#1})}
\newcommand{\rfn}[1]{(\ref{#1})}

\newcommand{\gsim}{\lower.7ex\hbox{$\;\stackrel{\textstyle>}{\sim}\;$}}
\newcommand{\lsim}{\lower.7ex\hbox{$\;\stackrel{\textstyle<}{\sim}\;$}}

\begin{document}

\title{Seesaw mechanism and the baryon asymmetry}

\author{M. RAIDAL}

\address{NICPB, Ravala 10, 10143 Tallinn, Estonia}

\maketitle

\abstracts{
I review the present understanding of connection between the
non-zero neutrino masses and the baryon asymmetry of the Universe.
The state-of-art results are presented for the standard thermal 
leptogenesis.
}

\section{Smallness of neutrino masses and the seesaw mechanism }

Non-zero  neutrino masses and mixing angles 
provide at the moment the only convincing  evidence~\cite{nuexp} of physics 
beyond the standard model. A paradigm to understand the smallness of
neutrino masses, involving some new heavy states which break lepton number, 
is called the seesaw mechanism~\cite{seesaw}. 
If the heavy particles couple to the Standard 
Model lepton and Higgs doublets, decoupling of them generates at low scale
the dimension five operator 
\bea
\frac{1}{\Lambda}LLHH.
\label{op}
\eea 
After the electroweak symmetry 
breaking \rfn{op} generates small neutrino masses suppressed 
by the heavy scale $\Lambda.$

According to the original proposal~\cite{seesaw}, which still is by far the
most popular and the most studied version of the seesaw mechanism, the
heavy states are three superheavy singlet Majorana neutrinos $N_i$. 
These can be identified with the right-handed chiral 
fields of some grand unification gauge group
such as $SO(10)$ and play a fundamental role in the anomaly cancellation.
As the singlets do not have the Standard Model gauge couplings, they do not 
spoil the nice features of the Standard Model or its supersymmetric extension,
such as the gauge coupling unification.
The relevant terms for the light neutrino masses in the Lagrangian 
are the neutrino Yukawa couplings $Y_\nu$ and Majorana masses $m_N,$ 
\beq
 L= \bar L_{Li} Y_\nu^{ij} N_{Rj} H + 
\frac{1}{2} \bar N_{Ri}^c m_N^{ij} N_{Rj} + 
\mrm{h.c.}
\eeq 
Integrating out the heavy neutrinos one obtains the seesaw relation for
light neutrino masses
\bea
m_\nu=- Y^T_\nu m^{-1}_N Y_\nu v^2,
\label{Nseesaw}
\eea
where $v=174$ GeV.
According to that, the relation 
between light neutrino masses and mixing in $m_\nu$, and the structure 
of the heavy neutrino Yukawa couplings and mass matrix is 
rather complicated, and obtaining the observed neutrino mixing pattern 
requires complicated flavour model building~\cite{AF}. 
As the deviation of the light neutrino mixing from bimaximal seems to be
parametrized by the quark mixing matrix, this quark-lepton complementarity
\cite{raidal} may indicate some unification effect for the fermion Yukawa
couplings. There are 9 physical degrees of freedom in the left-hand side
of \rfn{Nseesaw}  while the right-hand side contains 18 of them.
The missing 9 physical degrees of freedom at low energies 
can be parametrized by
an orthogonal parameter matrix $R$~\cite{ci} or by an Hermitian
parameter matrix $H$~\cite{di1}. The latter one allows to relate
the missing degrees of freedom to different observables in the 
supersymmetric models with universal boundary condition  for
the soft supersymmetry breaking terms \cite{Borzumati}.
Those include the renormalization induced 
lepton flavour violating processes \cite {Hisano} and electric dipole
moments~\cite{dipole}. The connection between low- and high-energy 
observables in the singlet seesaw models has been reviewed by
 S. Davidson in this conference \cite{Davidson}.

The second proposal for generating light neutrino masses is to couple the
Standard Model doublets to the $SU(2)_L$ triplet Higgs boson with non-zero
hypercharge~\cite{valle}. The relevant interaction terms are given by
\bea
L=\frac{1}{\sqrt{2}}(Y^{ij}_T \bar{L}_i^c i \tau_2 {\bf T} L_j + 
\lambda H^T {\bf T}^* i\tau_2 H + 
\mrm{h.c.})+ M_T \mrm{Tr}[{\bf T} {\bf T}^\dagger] ,
\eea 
where ${\bf T}=\tau \cdot T$,  the triplet $T$ is in the 
$SU(2)_L \times U(1)_Y$  representation $T\sim (3,1)$,
and the $\tau_i$ are the three Pauli matrices.  
Notice that the Yukawa couplings $Y_T$ and the 
Higgs self coupling $\lambda$ together break lepton number
explicitly. In this case the neutrino masses are suppressed
by the heavy triplet mass $M_T$ via the triplet seesaw 
mechanism~\cite{tseesaw}
\bea
m_\nu^{ij}=
Y_T^{ij} \lambda \frac{v^2}{M_T} .
\label{Tseesaw}
\eea
The triplet neutrino mass mechanism is very appealing
one from the low energy neutrino phenomenology point of view because
it requires introduction of the minimal number of new degrees of freedom
and because the neutrino masses are directly proportional (up to the
renormalization effects) to the triplet Higgs Yukawa couplings. In the
triplet seesaw mechanism the low energy neutrino mass measurements
determine directly, up to the overall scale, the structure of Majorana
type Yukawa couplings in the fundamental Lagrangian. Indeed, comparison of 
\rfn{Tseesaw} with \rfn{Nseesaw} shows that the flavour structure of
\rfn{Tseesaw} is trivial. Therefore the 
explanation to the almost bimaximal light neutrino mixing is free of
fine tunings and unnatural cancellations between numerical parameters.  
This simplicity also implies that the flavour violating processes in 
supersymmetric models are related to each other~\cite{anna}.
The scale of $M_T$ can vary from almost unification scale to 
as low as
1 TeV~\cite{mrs}. In the latter case  Higgs triplet can be discovered
in the future collider experiments~\cite{lhc}.

The third neutrino mass mechanism with triplet fermions~\cite{ma}
is considered to be somewhat exotic and has not gained much attention
in the neutrino model building industry. We do not discuss this
possibility in this talk any further.

\section{ The seesaw mechanism and leptogenesis}

Another observable related to the physics of neutrino masses is the
baryon asymmetry of the Universe. The observed ratio of baryon density to 
entrophy density~\cite{wmap},
\bea
\frac{n_B}{s}=(8.7\pm 0.4) \times 10^{-11},
\label{ns}
\eea
requires the existence of physics beyond the Standard Model. 
To generate \rfn{ns}, three famous Sakharov's conditions must be satisfied
\cite{Sakharov}.
Firstly, baryon number must be violated. Secondly, C and CP must be
violated. Thirdly, the process must take place in out-of-equilibrium 
situation. In principle, those conditions can be satisfied also in the 
Standard Model since at non-perturbative level both $B$ and $L$ 
are separately violated~\cite{thooft}. However, 
baryogenesis in the electroweak phase transition has been extensively 
studied and found not to able to generate \rfn{ns} because it requires 
very light Higgs boson mass $M_H<40$ GeV. Therefore,  
currently the most widely accepted concept for generating \rfn{ns} 
is leptogenesis. 

According to the paradigm of leptogenesis~\cite{fy}, non-zero lepton asymmetry 
is generated first in out-of-equilibrium decays of some heavy states in the
early universe. Thus the interactions of those heavy particles must
violate lepton number and CP, and out-of-equilibrium condition is
provided by the expansion of the Universe, $\Gamma<H$, where $H$ is the
Hubble parameter. Thereafter the lepton asymmetry is reprocessed to the
$B-L$ asymmetry by the sphaleron processes~\cite{Kuzmin}, generating \rfn{ns}.
Baryogenesis via leptogenesis is the only idea which is also supported
by the experimental data, namely by the non-vanishing neutrino masses
and mixing. If we require the seesaw mechanism to induce simultaneously 
both the neutrino masses and the baryon asymmetry of the Universe,
one can constrain leptogenesis from the experimental neutrino data.
From that point of view different realizations of the seesaw
mechanism discussed in the previous Section have quite different
features. 
In this talk I consider only thermal leptogenesis, \ie,
the case when all the particle species are created by thermal
plasma during and after reheating of the Universe. All the following discussion
applies only to that case.

In the case of the singlet neutrino seesaw mechanism~\cite{seesaw}
leptogenesis~\cite{fy} is a direct, almost un-avoidable, consequence 
of seesaw rather that a separate mechanism.  
Because present neutrino data requires
the existence of at least two heavy singlet neutrinos, and because in the
case of Majorana particles physical CP phases exist already in the
case of two generation, even the most minimal singlet seesaw model 
implies viable leptogenesis~\cite{fgy}. If the heavy and light neutrinos are 
hierarchical in mass so that leptogenesis comes from the decays of the
lightest singlet $N_1$ only, there exist an upper bound on the CP asymmetry
from its decay~\cite{di2}
\bea
\epsilon_{N_1}
\le \frac{3}{16\pi}
\frac{m_{N_1} m_3}{v^2},
\label{epsbound}
\eea
 where $m_3$ is the heaviest light neutrino mass. 
In more general case there is an upper bound on the the light neutrino mass
scale from leptogenesis~\cite{nubound} which does not allow highly degenerate
neutrino masses. This, together with
\rfn{ns} and with the state-of-art estimates of the thermal washout 
effects~\cite{gnrrs}, 
leads to the lower bound on the leptogenesis scale, which turns out
to be $m_{N_1}>2\times 10^9$ GeV. 
However, if the singlet neutrinos
are partially degenerate in mass, the CP asymmetry is resonantly enhanced
\cite{reson,crv} and leptogenesis scale as low as ${O}(1)$ TeV 
could be viable. In principle the bound \rfn{epsbound} can also be violated
if the heavy neutrinos are not degenerate~\cite{tfermion}.
First neutrino model of that sort has been proposed in Ref.~\cite{rst},
and such models are quite different from the generic ones.
All-together, leptogenesis is a very natural
consequence of the singlet seesaw mechanism.

On the other hand, the triplet seesaw mechanism, which in its minimal
form contains just one triplet Higgs, does not
provide leptogenesis in the Standard Model because of the lack of interfering
amplitudes. The minimal triplet leptogenesis model must contain two
triplets~\cite{ms} which doubles the neutrino degrees of freedom, and
 the nicest phenomenological argument of simplicity in favour of this scenario
is lost. However, in the minimal supersymmetric version of the
triplet seesaw model leptogenesis is possible~\cite{tsoft}.
In the supersymmetric triplet seesaw model the anomaly cancellation
requires introduction of two triplets with opposite $U(1)$ quantum
numbers, $T\sim (3,1)$ and $\bar T\sim (3,-1)$. According to the 
superpotential
\bea
W\!= \!
\frac{1}{\sqrt{2}}(Y^{ij}_T L_i T L_j + 
\lambda_1 H_1 T H_1 + \lambda_2 H_2\bar T H_2)+ M_T T \bar T , 
\label{W}
\eea
$T$ and $\bar T$ have equal masses but only one of them couples to the lepton
doublet, thus giving \rfn{Tseesaw} with $\lambda=\lambda_2,$ $v=v_2.$
When supersymmetry breaking terms are included, $T$ and $\bar T$
degeneracy is split by the soft terms, and resonant leptogenesis occurs 
\cite{tsoft} via the mechanism called ``soft leptogenesis''~\cite{soft}. 
While the scale of triplet seesaw is rather arbitrary, the scale of
triplet leptogenesis is very much constrained due to the strong
dependence on the size of the soft supersymmetry breaking
$B$ terms.
In the following discussion we concentrate only on the singlet
leptogenesis as the most interesting one.

\section{Standard thermal leptogenesis}

By the standard leptogenesis we mean the evolution of lepton asymmetry 
in the thermal plasma during and after reheating of the Universe. 
The heavy neutrinos are assumed to be hierarchical in mass
so that leptogenesis occurs in the decays of the lightest heavy neutrinos
$N_1$ only. Despite of the fact that the neutrino sector of
the minimal seesaw model contains 18 free parameters, this scenario 
depends on only very few combinations of them. We parametrize 
the generated $B-L$ asymmetry via
\begin{equation}
Y_{{{B}-{ L}}}=-\epsilon_{N_1} ~ \eta ~ Y^{\rm eq}_{N_1}
(T\gg m_{N_1}),
\label{b-l}
\end{equation}
where $\epsilon_{N_1}$ is the CP-asymmetry in $N_1$
decays at zero temperature given by~\cite{crv} 
\begin{equation}
 \epsilon_{N_1}(T=0)=\frac{1}{8\pi}\sum_{j\neq 1}
\frac{\textrm{Im}
\left[ (Y^{\dagger} Y)_{j 1}^2\right] }{\left[Y^{\dagger}
Y\right]_{11}}
f\left(\frac{m_{N_j}^2}{m_{N_1}^2}\right),
\label{eps0}
\end{equation}
with
\begin{equation}
f(x)=\sqrt{x}\left[ \frac{x-2}{x-1}-(1+x)\ln
\left(\frac{1+x}{x} \right)
 \right] \stackrel{x\gg 1}{\longrightarrow} - \frac{3}{2
\sqrt{x}} \ ,
\label{f(x)}
\end{equation}
and $Y^{\rm eq}_{N_1}(T\gg
m_{N_1}) = 135 \zeta(3)/(4\pi^4g_*)$ is the neutrino equilibrium
density at high temperature. Here $g_*$ counts the effective
number of spin-degrees of freedom in thermal equilibrium 
($g_*= 106.75$ in the SM with no right-handed neutrinos).
If the light (and also heavy) neutrino masses are taken to be 
hierarchical,
the lower bound \rfn{epsbound} on the CP asymmetry holds. 
In the following we assume that this is the case. 
\Eq{b-l} can be regarded as the definition of the leptogenesis
efficiency parameter $\eta$, which contains all the  finite
temperature effects and dependences on the initial conditions for 
heavy neutrino abundances. For hierarchical neutrino masses, and for 
particular initial condition, the efficiency $\eta$ depends on
only two parameters. Those are $m_{N_1}$ and the 
effective neutrino mass \cite{}
\bea
\tilde m_1=\frac{(Y_\nu Y_\nu^\dagger)_{11}}{m_{N_1}} v^2,
\eea
which is the measure on $N_1$ interaction strength with thermal
plasma.

The obtained $B-L$ asymmetry is converted to the
baryon asymmetry via sphaleron processes
\begin{equation}\label{eq:YB}
\frac{n_{B}}{s}=\frac{24+4n_H}{66+13n_H}\frac{n_{{ B} - { L}}}{s},
\end{equation}
where $n_H$ is the number of Higgs doublets.  For the SM we
find numerically
\begin{equation}
\label{uusm}
\frac{n_{ B}}{s} =-1.38\times 10^{-3} \epsilon_{N_1} \eta.
\end{equation}

\begin{figure}[t]
\centerline{\epsfxsize = 0.5\textwidth \epsffile{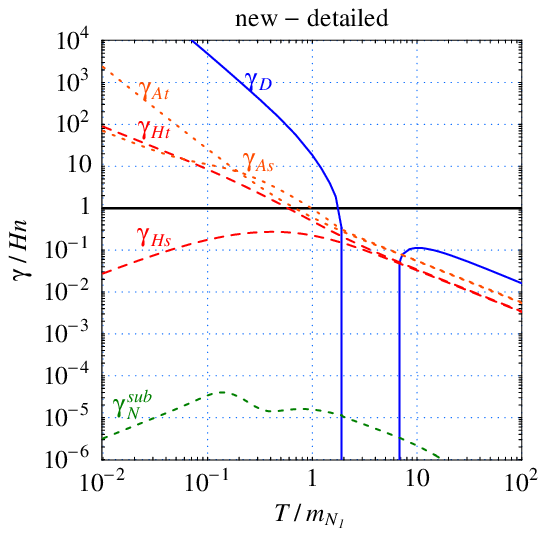}
\hfill \epsfxsize = 0.5\textwidth \epsffile{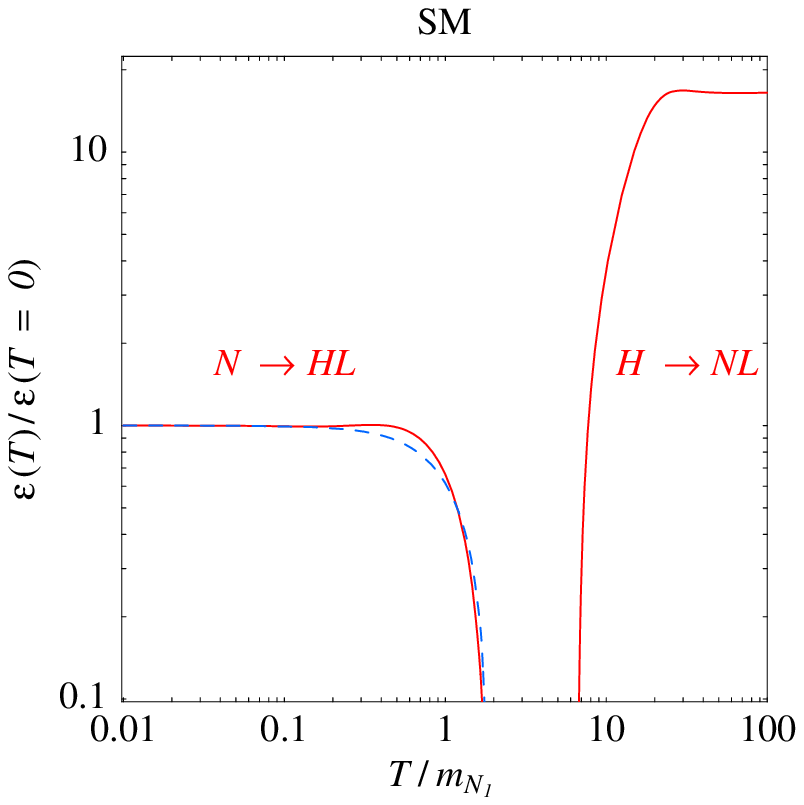}
}
\caption{
Evolution of the scattering densities (first plot) and the
CP asymmetries (second plot) with temperature. 
\label{ampl}}
\end{figure}

Although the standard thermal leptogenesis scenario has been
studied extensively \cite{pluemi}, several important aspects
of this scenario have been worked out in detail quite recently. 
Most important of them is adding finite temperature 
corrections to the decay and scattering amplitudes and to the CP 
asymmetry \cite{gnrrs}. It has been found that to a good approximation
the dominant effects are included if one (i) uses thermal
masses of particles in the processes involved 
instead of zero temperature masses; (ii)
renormalizes all the 
couplings at the first Matsubara mode,
\begin{equation}
E_{r}=2 \pi T,
\label{scale}
\end{equation}
using the zero-temperature renormalization group equations;
(iii) uses finite temperature Feynman rules for calculation of the 
CP asymmetries. 
The important consequence of this is that at high temperatures
the Higgs boson mass exceeds the sum of the singlet neutrino and 
lepton mass, and instead of $N$ decay the two-body decay
process is $H\to N L.$
The second important refinement of the calculation is inclusion of
the gauge scatterings in \cite{gnrrs} (which in particular
limit  agree with the similar attempt in \cite{pu}).
Thus the processes contributing to the thermal leptogenesis
are
\beq\begin{array}{lll}
\Delta L = 1:\qquad& D = [N_1\leftrightarrow LH], &
S_s = H_s + A_s, \qquad
S_t= H_t + A_t, \\[3mm]
\Delta L = 2: & N_s=[LH \leftrightarrow\bar{L}\bar{H}],&
\qquad N_t=[LL \leftrightarrow\bar{H}\bar{H}],
\end{array}
\eeq
where
\beq\begin{array}{ ll}
H_s = [LN_1 \leftrightarrow Q_3U_3],      &  
2H_t =[N_1\bar U_3 \leftrightarrow Q_3\bar L]+  
[N_1\bar Q_3\leftrightarrow U_3\bar L],  \\
A_s=[LN_1\leftrightarrow \bar H A],      &   
2A_t=  [N_1 H \leftrightarrow A {\bar L}]+  [N_1 A\leftrightarrow {\bar H}
{\bar L}].
\end{array}
\eeq
We separated $\Delta L = 1$ scatterings $S_{s,t}$ into top ($H_{s,t}$) and
gauge contributions ($A_{s,t}$).
The evolution of scattering densities $\gamma$ for the particular processes
with temperature has been shown in Fig. \ref{ampl}  for 
$m_{N_1}=10^{10}$ GeV and $\tilde m_1=0.06$ eV.
It follows that there is a temperature range in which $H$ and $L$ thermal
masses are such that all the two-body decays are kinematically forbidden.
Notice that in order to avoid double counting of the two-body decays,
after subtracting the resonances from the scattering
$LH\leftrightarrow LH$ the corresponding $\gamma_N^{sub}$
is completely negligible compared to the other processes.
In the same figure we also plot the dependence of the CP asymmetries in
$N\to HL$ and in $H\to NL$ decays. The solid line shows the result
of full finite temperature calculation in \cite{gnrrs} while the
dashed line is obtained by approximating $N_1$ to be at rest
in thermal plasma. Those results agree with each other with good accuracy.
The results including thermal corrections plotted in Fig. \ref{ampl}
differ both qualitatively and quantitatively from the zero temperature
calculations and change the predictions for the generated $B-L$ asymmetry.

\begin{figure}[t]
\centerline{\epsfxsize = 0.5\textwidth \epsffile{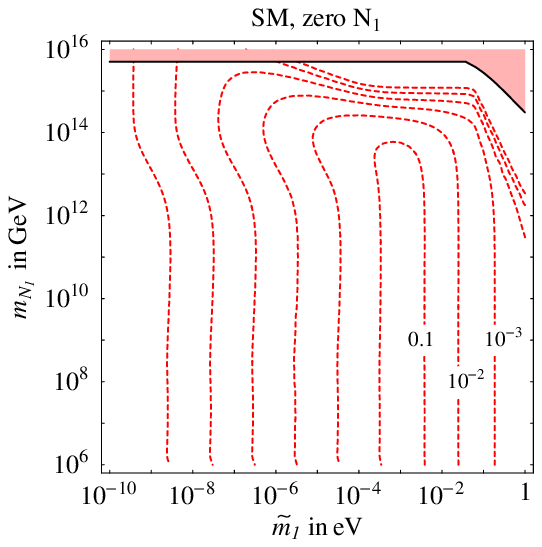}
\hfill \epsfxsize = 0.5\textwidth \epsffile{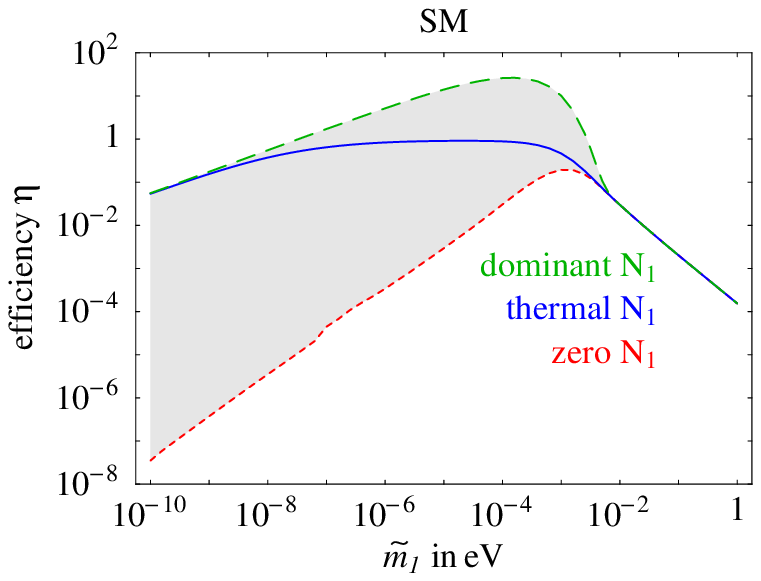}
}
\caption{
Isocurves of leptogenesis efficiency $\eta$. The first plot
is for vanishing initial $N_1$ abundance, the second plot is for fixed
$m_{N_1}=10^{10}$ GeV. 
\label{teffect}}
\end{figure}

Solving the full set of Boltzmann equations for the evolution of 
the $B-L$ asymmetry with temperature,  and 
 parameterizing the results via \Eq{b-l}, our results for the 
leptogenesis efficiency $\eta$ are shown in  Fig. \ref{teffect}.
In the first plot we present the isocurves of $\eta=10^i,$ $i=-1,-2,-3,...$
in the $(\tilde m_1, m_{N_1})$ plane, assuming vanishing initial
abundance for the singlet neutrinos $Y_{N_1}(T=\infty)=0.$
The maximum efficiency is found around $\tilde m_1\sim 10^{-3}$ eV.
For smaller values of $\tilde m_1$ the Yukawa interactions of $N_1$ 
are too weak for $N_1$ to reach thermal abundance, which leads to
the suppression of the $B-L$ asymmetry. For larger values of $\tilde m_1$ 
the Yukawa interaction becomes too strong so that the $N_1$ decays
occur not sufficiently in out-of-equilibrium,  which again leads to
the suppression of the $B-L$ asymmetry.
Our result can be summarized with a simple analytical fit \cite{gnrrs}
\begin{equation}
\frac{1}{\eta}\approx  \frac{3.3\times 10^{-3}\mrm{eV}}{\tilde{m}_1} +   
\bigg(\frac{\tilde{m}_1}{0.55\times 10^{-3}\mrm{eV}}\bigg)^{1.16},
\end{equation}
valid for $m_{N_1}\ll 10^{14}$ GeV. 
This enables the  reader to study leptogenesis 
in neutrino mass models without setting up and solving the
complicated Boltzmann equations. 

The dependence of the efficiency on the initial conditions for $N_1$
is demonstrated on the second plot in Fig. \ref{teffect} where we assume 
$m_{N_1}=10^{10}$ GeV and  
study three different cases, zero initial abundance for $N_1$, 
thermal initial abundance for $N_1$ and when $N_1$ dominates the Universe.
For small values of $\tilde m_1$ the results depend on the initial
conditions. However, if $\tilde m_1=\sqrt{\Delta m_{sol}^2}$ or  
$\tilde m_1=\sqrt{\Delta m_{atm}^2}$, the prediction of thermal
leptogenesis is practically independent of any pre-existing 
initial condition. This makes the thermal leptogenesis
predictions so robust.

In order to study in which parameter space the thermal
leptogenesis is capable to yield the observed baryon
asymmetry of the Universe, one has to combine the results
for the efficiency $\eta$ with the maximal value of the
CP asymmetry (corrections to \Eq{epsbound} from non-zero $m_1$ 
have to be taken into account).
\begin{figure}[t]
\centerline{\epsfxsize = 0.5\textwidth \epsffile{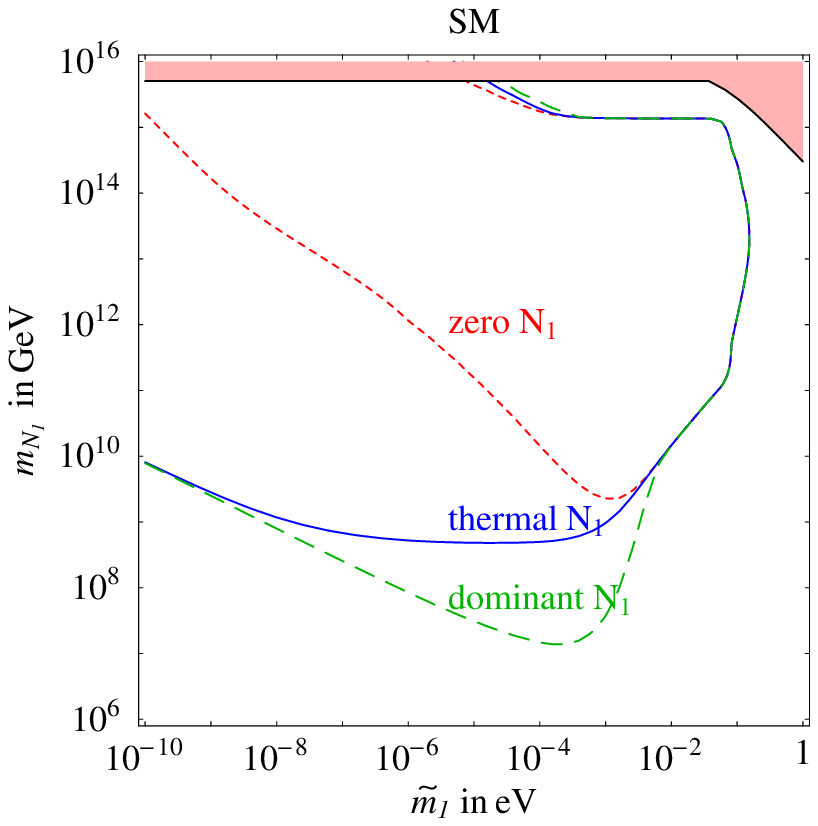}
\hfill \epsfxsize = 0.5\textwidth \epsffile{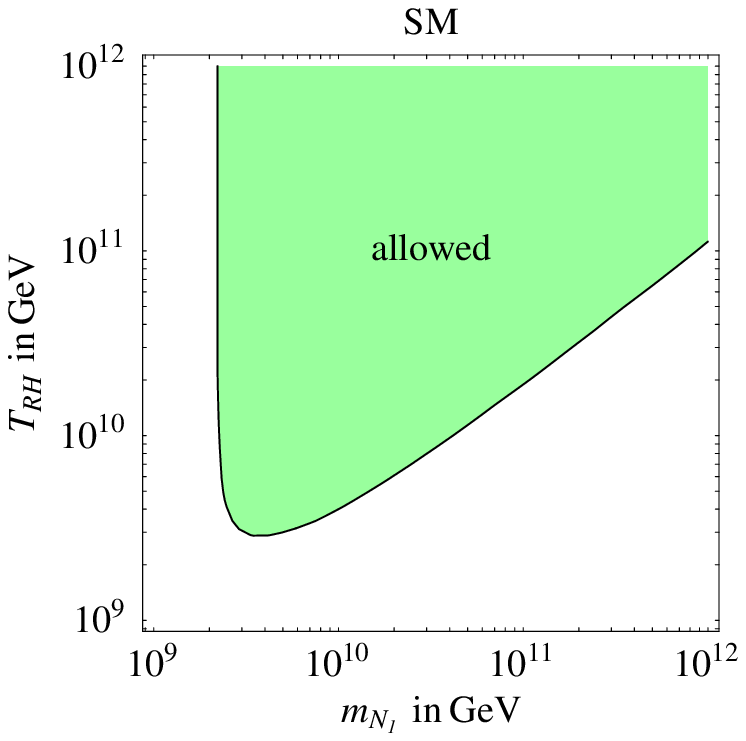}
}
\caption{Regions in $(\tilde m_1, m_{N_1})$ and 
$(m_{N_1}, T_{RH})$ planes allowed by successful leptogenesis. 
\label{lower}}
\end{figure}
This has been done in Fig. \ref{lower}. In the first plot in this
figure we present the contours for successful leptogenesis in the 
 $(\tilde m_1, m_{N_1})$ plane for the three initial conditions for
$N_1$ as before. There exist lower bounds on the $N_1$ mass from the
requirement of successful leptogenesis which depend on the initial
conditions. For the vanishing initial $N_1$ abundance, which is 
probably the most appropriate assumption for singlets, the
bound is \cite{gnrrs}
\bea
m_{N_1}>2 \times 10^9 \,\,\,\, \mrm{GeV}.
\eea

To derive the results presented so far we have assumed that the reheating 
temperature of the Universe exceeds the singlet neutrino mass, 
$T_{RH}\gg m_{N_1}.$ This may not be the case in all the scenarios, 
especially in supersymmetric ones because supersymmetry sets an
upper bound on the reheating temperature of the Universe from the
overproduction of gravitinos \cite{gr}. Therefore one has to study how 
our predictions depend on the reheating of the Universe. We have set up and 
solved Boltzmann equations which include $T_{RH}$ as a free
parameter in Ref. \cite{gnrrs}. The result is presented 
on the second plot in Fig. \ref{lower} where we show the
region in the $(m_{N_1}, T_{RH})$ plane allowed by thermal leptogenesis.
We have taken $\tilde m_1=10^{-3}$ eV and assumed vanishing initial
abundance for $N_1.$ The results show that $T_{RH}$ can be as low as
$m_{N_1}$ without considerable loss of the efficiency.
However, if $T_{RH} < m_{N_1}$ successful leptogenesis  requires 
large values of $m_{N_1},$ considerably larger than the minimally
allowed one.

\section{Conclusions}

Baryogenesis via leptogenesis, the idea of generating the
observed baryon asymmetry of the Universe in out-of-equilibrium 
decays of heavy particles which violate lepton number, 
is supported by the light neutrino data.
Remarkably, the standard thermal leptogenesis works in particularly
robust way in the neutrino parameter space suggested by the seesaw
mechanism. Currently it is not possible to make exact predictions
between the low energy neutrino measurements and the generated
baryon asymmetry  because of too many free parameter in the seesaw model.
Nevertheless the experimental and theoretical success in 
understanding the seesaw meshanism during its first 25 years 
has been impressive. 
Hopefully in next 25 years, if next generation experiments will find new  
lepton flavour violating observables, leptogenesis can be
experimantally established as the mechanism
of baryogenesis.

\section*{Acknowledgments}
I would like to thank all my collaborators with whom I have studied
the issues presented in this talk over several past years.
This work has been supported by the ESF Grants 5135 and 5935, by the
EC MC contract MERG-CT-2003-503626, and by the Ministry of Education and 
Research of the Republic of Estonia.

\end{document}